\documentclass[conference]{IEEEtran}

\usepackage{times}
\usepackage[final]{graphicx}
\usepackage[reqno]{amsmath}
\usepackage{amsfonts}
\usepackage{times,amsmath,epsfig}
\usepackage{latexsym,amssymb}
\usepackage{cite}

\begin{document}

\title{Blind Cognitive MAC Protocols}

\author{\IEEEauthorblockN{Omar Mehanna, Ahmed Sultan}
\IEEEauthorblockA{Wireless Intelligent Networks\\ Center  (WINC)\\Nile University, Cairo, Egypt\\omar.mehanna@nileu.edu.eg, asultan@nileuniversity.edu.eg}

\and
\IEEEauthorblockN{Hesham El Gamal}
\IEEEauthorblockA{Department of Electrical \\and Computer Engineering\\Ohio State University, Columbus, USA\\helgamal@ece.osu.edu}}

\maketitle

\begin{abstract}
We consider the design of cognitive Medium Access Control (MAC)
protocols enabling an unlicensed (secondary) transmitter-receiver
pair to communicate over the idle periods of a set of licensed
channels, i.e., the primary network. The objective is to maximize
data throughput while maintaining the synchronization between
secondary users and avoiding interference with licensed (primary)
users. No statistical information about the primary traffic is
assumed to be available {\em a-priori} to the secondary user. We
investigate two distinct sensing scenarios. In the first, the
secondary transmitter is capable of sensing all the primary
channels, whereas it senses one channel only in the second
scenario. In both cases, we propose MAC protocols that efficiently
learn the statistics of the primary traffic on-line. Our
simulation results demonstrate that the proposed blind protocols
asymptotically achieve the throughput obtained when prior
knowledge of primary traffic statistics is available.

\end{abstract}

\section{Introduction}
Most of licensed spectrum resources are under-utilized. This
observation has encouraged the emergence of dynamic and
opportunistic spectrum access concepts, where unlicensed (secondary)
users equipped with cognitive radios are allowed to
opportunistically access the spectrum as long as they do not
interfere with licensed (primary) users. To achieve this goal,
secondary users must monitor the primary traffic in order to
identify spectrum holes or opportunities which can be exploited to
transfer data~\cite{Haykin}.

The main goal of a cognitive MAC protocol is
to sense the radio spectrum, detect the occupancy state of different
primary spectrum channels, and then opportunistically communicate
over unused channels (spectrum holes) with minimal interference to
the primary users. Specifically, the cognitive MAC protocol should
continuously make efficient decisions on which channels to sense and
access in order to obtain the most benefit from the available
spectrum opportunities. Several cognitive MAC protocols have been
proposed in previous studies. For example, in~\cite{HangSu}, MAC
protocols were constructed assuming each secondary user is equipped
with two transceivers, a control transceiver tuned to a dedicated
control channel and a software defined radio SDR-based transceiver
tuned to any available channels to sense, receive, and transmit
signals/packets. On the other hand,~\cite{HyoilKim} proposed a
sensing-period optimization mechanism and an optimal
channel-sequencing algorithm, as well as an environment adaptive
channel-usage pattern estimation method.

The slotted Markovian structure for the primary network traffic,
adopted here, was also considered in~\cite{Zhao} where the optimal
policy was characterized and a simple greedy policy for secondary
users was constructed. The authors of~\cite{Zhao}, however,
assumed that the primary traffic statistics (i.e., Markov chain
transition probabilities) were available {\em a-priori} to the
secondary users. Here, our focus is on the blind scenario where
the cognitive MAC protocol must learn the transition probabilities
on-line.

In this work, we differentiate between two scenarios. The first
assumes that the secondary transmitter can sense all the available
primary channels before making the decision on which one to
access. The secondary receiver, however, does not participate in
the sensing process and can {\em wait to decode} on only one
channel. This is the model adopted in~\cite{Capacity}. In the
sequel, we propose an efficient algorithm that optimizes the
on-line learning capabilities of the secondary transmitter and
ensures perfect synchronization between the secondary pair. The
proposed protocol does not assume a separate control channel, and
hence, piggybacks the synchronization information on the same data
packet. Our numerical results demonstrate the superiority of the
proposed protocol over the one in~\cite{Capacity} where the
primary transmitter and receiver are assumed to access the channel
in a predetermined sequence, which they agreed upon {\em
a-priori}.

The second scenario assumes that both the secondary transmitter and
receiver can sense only one primary channel in each time slot. This
problem can be re-casted as a restless multi-armed bandit problem
where the optimal algorithm must strike a balance between
exploration and exploitation~\cite{ZhaoWhitle}. Unfortunately,
finding the optimal solution for this problem remains an elusive
task~\cite{Whittle}. Inspired by the recent results
of~\cite{ZhaoWhitle} and~\cite{Lifeng}, an efficient MAC protocol is
constructed which can be viewed as the Whittle index strategy
of~\cite{ZhaoWhitle} augmented with a similar learning phase to the
one proposed in~\cite{Lifeng} for the multi-armed bandit scenario.
Our numerical results show that the performance of this protocol
converges to the Whittle index strategy with known transition
probabilities~\cite{ZhaoWhitle}.

\section{Network Model}

\subsection{Primary Network}
We consider a primary network consisting of $N$ independent
channels with its users communicating according to a synchronous
slot structure. We use $i$ to refer to the channel index
$i\in\{1,\cdots,N\}$, and $j$ to refer to the time-slot index
$j\in\{1,\cdots,T\}$. The {\it i}th primary channel has a
bandwidth of $B_i$. The traffic statistics of the primary network
are such that the occupancy of each of the $N$ channels follows a
discrete-time Markov process with two states. The state of the
{\it i}th channel at time slot $j$, $S_i^{(j)}$, is equal to $1$
if the channel is free, and to $0$ if it is busy. The state
diagram for a single Markov channel model is illustrated in Figure
1. The channel state transition matrix of the Markov chain is
given by $P^i = \left[
   \begin{array}{cc}
     P_{00}^i & P_{01}^i \\
     P_{10}^i & P_{11}^i \\
   \end{array}
 \right]$.
We assume that $P^i$ remains fixed for a block of $T$ time slots
and is unknown {\em a-priori} to the secondary user.

\begin{figure}
  \includegraphics[width=0.5 \textwidth]{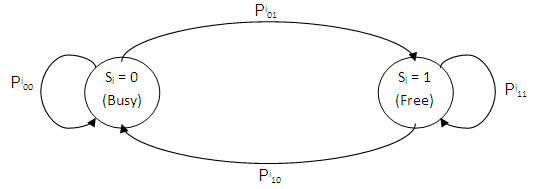}
  \caption{The Gilber-Elliot channel model}\label{ChanState}
\end{figure}

\subsection{Secondary Pair}

It is assumed that the secondary transmitter can sense $L_1$
channels $(L_1 \leq N)$ and can access $L_2=1$ channel in each slot.
The secondary transmitter can only transmit if the channel it
chooses to access is sensed to be free. Here, we only report our
results for the two special cases $L_1=N$ and $L_1=1$. The more
general case will be addressed in the journal version.

The secondary receiver does not participate in channel sensing and
is assumed to be capable of accessing only one
channel~\cite{Capacity}. This assumption is intended to limit the
decoding complexity needed by the secondary receiver. Another
motivation behind restricting channel sensing to the transmitter
is the potentially different sensing outcomes at the secondary
transmitter and receiver due to the spatial diversity of the
primary traffic which can lead to the breakdown of the secondary
transmitter-receiver synchronization.

Conceptually, our proposed cognitive MAC protocol can be
decomposed into the following stages:
\begin{itemize}
    \item \textit{Decision stage}: The secondary transmitter decides which $L_1$ channels to sense. Also,
    both transmitter and receiver decide which channel to access.
    \item \textit{Sensing stage}: The transmitter senses the $L_1$ selected primary channels.
    \item \textit{Learning stage}: The transmitter updates the estimated primary channels' statistics,
    $\hat{P^i}$.
    \item \textit{Access stage}: If the access channel is sensed to be free, a data packet is transmitted to the secondary
    receiver. This packet contains the information needed to sustain synchronization between secondary
    terminals and, hence, synchronization does not require a dedicated control channel. The length of the packet is assumed to be large
    enough such that the loss of throughput resulting from the
    synchronization overhead is marginal.
    \item \textit{ACK stage}: The receiver sends an ACK to the transmitter upon successful reception of sent data.
\end{itemize}

The performance of the sensing stage is limited by two types of
errors. If the secondary transmitter decides that an empty channel
is busy, it will refrain from transmitting, and a spectrum
opportunity is overlooked. This is the false alarm situation,
which is characterized by probability of false alarm $P_{FA}$. On
the other hand, if the detector fails to sense a busy channel as
busy, a miss detection occurs resulting in interference with
primary user. The probability of miss detection is denoted by
$P_{MD}$. In the rest of the paper, $\bar{S}_i^{(j)}$ denotes the
state of channel $i$ at time slot $j$ as sensed by the
transmitter, which might not be the actual channel state
$S_i^{(j)}$. Overall, successful communication between the
secondary transmitter and receiver occur only when: 1) they both
decide to access the same channel, and 2) the channel is sensed to
be free and is actually free from primary transmissions.

\section{Full Sensing Capability: $L_1 = N$}

In this section it is assumed that the secondary transmitter can
sense all $N$ primary channels at the beginning of each time slot.
The initial packet sent to the receiver includes estimates for the
transition probabilities, and the belief vector
$\bar{\Omega}^{(1)}$, where $ \bar{\Omega}^{(j)} =
[\bar{\omega}_1^{(j)},\cdots,\bar{\omega}_N^{(j)}]$, and
$\bar{\omega}_i^{(j)}$ is the {\em common} transmitter's and
receiver's estimate of the prior probability that channel $i$ is
free at the beginning of time slot $j$, on the basis of the
sensing history of channel $i$. Once the initial communication is
established, the secondary transmitter and receiver implement the
same spectrum access strategy described below for $j \geq 1$.

\begin{enumerate}
  \item \emph{Decision:} At the beginning of time slot $j$, and using belief vector $\bar{\Omega}^{(j)}$, the secondary transmitter and receiver decide to access channel\\
      $i^*(j) = \arg \max\limits_{i = 1,\cdots, N} \left[\bar{\omega}_i^{(j)}B_i\right] $
  \item  \emph{Sensing:} The secondary transmitter senses all channels and captures the sensing vector
        $\Phi^{(j)} = [\bar{S}_1^{(j)},\cdots,\bar{S}_N^{(j)}]$, where $\bar{S}_i^{(j)}=1$
        if the {\it i}th channel is sensed to be free, and $\bar{S}_i^{(j)}=0$ if it is found busy.
  \item \emph{Learning:} Based on the sensing results, the transmitter updates the estimates $\hat{P}_{01}^i$ and $\hat{P}_{11}^i$ for all primary
  channels as explained below.
  \item  \emph{Access:} If $\bar{S}_{i^*}^{(j)} = 1$, the transmitter sends its data packet to the receiver. The packet includes $\Phi^{(j)}$, $\hat{P}_{01}^i$ and $\hat{P}_{11}^i$.
      In addition, if the transmission at slot $j-1$ has failed, the transmitter sends $\Omega^{(j)}$, which is the belief vector computed at the transmitter based on its observations.
      If the receiver successfully receives the packet, it sends an ACK back to the
      transmitter. Parameter $K_{i^*}^{(j)}$ is equal to unity if an
      ACK is received by the transmitter, and zero otherwise. If the
      channel is free, the forward transmission and the feedback
      channel are assumed to be error-free.
  \item Finally, the transmitter and receiver update the common belief vector
  $\bar{\Omega}^{(j+1)}$ such that:
\end{enumerate}
\footnotesize
\begin{equation}\label{1}
    \bar{\omega}_i^{(j+1)} = \begin{cases}
   \bar{P}_{11}^i & \text{if $K_{i^*}^{(j)}= 1, i = i^*(j)$}\\
   \bar{A}_i\bar{P}_{11}^i +  \left(1-\bar{A}_i\right)\bar{P}_{01}^i & \text{if $K_{i^*}^{(j)}= 1, i \neq i^*(j), \bar{S}_i^{(j)} = 1$}\\
   \bar{C}_i\bar{P}_{11}^i +  \left(1-\bar{C}_i\right)\bar{P}_{01}^i   & \text{if $K_{i^*}^{(j)}= 1, i \neq i^*(j), \bar{S}_i^{(j)} = 0$}\\
   \bar{D}_i\bar{P}_{11}^i +  \left(1-\bar{D}_i\right)\bar{P}_{01}^i & \text{if $K_{i^*}^{(j)} = 0, i = i^*(j)$} \\
    \bar{\omega}_i^{(j)}\bar{P}_{11}^i + (1-\bar{\omega}_i^{(j)})\bar{P}_{01}^i  & \text{if $K_{i^*}^{(j)}=0, i \neq i^*(j)$}
    \end{cases}
\end{equation}

\normalsize
\noindent where:\\
$\bar{A}_i = Pr(S_i^{(j)} = 1 | \bar{S}_i^{(j)} = 1) = \frac{(1-P_{FA})\bar{\omega}_{i}^{(j)}}{(1-P_{FA})\bar{\omega}_{i}^{(j)} + P_{MD}(1-\bar{\omega}_{i}^{(j)})}$,\\
$\bar{C}_i = Pr(S_i^{(j)} = 1 | \bar{S}_i^{(j)} = 0) = \frac{P_{FA}\bar{\omega}_{i}^{(j)}}{P_{FA}\bar{\omega}_{i}^{(j)} + (1-P_{MD})(1-\bar{\omega}_{i}^{(j)})}$,\\
$\bar{D}_i = Pr(S_{i^*}^{(j)} = 1 | K_{i^*}^{(j)} = 0) = \frac{P_{FA}\bar{\omega}_{i}^{(j)}}{P_{FA}\bar{\omega}_{i}^{(j)} + (1 - \bar{\omega}_{i}^{(j)})}$,\\
$\bar{P}_{01}^i$ and $\bar{P}_{11}^i$ are the most recent shared
estimates of {\it i}th channel transition probabilities.
Obviously, in case of perfect sensing, $\bar{A}_i=1$,
$\bar{C}_i=0$ and $\bar{D}_i=0$.

In addition, the transmitter computes another belief vector,
$\Omega^{(j+1)}$, based on its observations: \footnotesize
\begin{equation}\label{2}
\omega_i^{(j+1)} =
\begin{cases}
   \bar{\omega}_i^{(j+1)} & \text{if $K_{i^*}^{(j)}= 1$}\\
   A_i\hat{P}_{11}^i +  \left(1-A_i\right)\hat{P}_{01}^i & \text{if $K_{i^*}^{(j)}= 0, i \neq i^*(j), \bar{S}_i^{(j)} = 1$}\\
   C_i\hat{P}_{11}^i +  \left(1-C_i\right)\hat{P}_{01}^i   & \text{if $K_{i^*}^{(j)}= 0, i \neq i^*(j), \bar{S}_i^{(j)} = 0$}\\
   D_i\hat{P}_{11}^i +  \left(1-D_i\right)\hat{P}_{01}^i & \text{if $K_{i^*}^{(j)} = 0, i = i^*(j)$}
\end{cases}
\end{equation}
\normalsize where $A_i$, $C_i$, and $D_i$ are the same as
$\bar{A}_i$, $\bar{C}_i$ and $\bar{D}_i$ with
$\bar{\omega}_i^{(j)}$ replaced by $\omega_i^{(j)}$. Note that
$\bar{\Omega}^{(1)}=\Omega^{(1)}$, and $\Omega^{(j+1)}$ differs
from $\bar{\Omega}^{(j+1)}$ only when $K_{i^*}^{(j)}= 0$. If
transmission succeeds at the {\it j}th time slot after one or more
failures, the transmitter and receiver set
$\bar{\Omega}^{(j)}=\Omega^{(j)}$ before computing
$\bar{\Omega}^{(j+1)}$.

Since we assume that traffic statistics on primary channels
($P^i$) are unknown to the secondary users {\em a-priori}, the
secondary users need to estimate these probabilities. When
continuous observations of each channel are available, each
channel can be modeled as a hidden Markov model (HMM). An optimal
learning algorithm for HMM is described in~\cite{HMM} using which
the transition probabilities, $P_{FA}$, and $P_{MD}$ can be
estimated. However, we propose a much less complex algorithm based
on simple counting, which approximates the estimated probabilities
by the optimal HMM algorithm. The algorithm we propose works as
follows. After sensing all the primary channels at the beginning
of each time slot, the secondary transmitter keeps track of the
following metrics for each channel:

\begin{itemize}
\item Number of times each channel was sensed to be free:\\
$N_1^i(j) = \sum\limits_{l=1}^{j-1} \bar{S}_i^{(l)}$ \item Number
of times each channel was sensed to be busy:\\ $N_0^i(j) =
\sum\limits_{l=1}^{j-1} (1-\bar{S}_i^{(l)})$ \item Number of state
transitions from free to free:\\ $N_{11}^i(j) =
\sum\limits_{l=1}^{j-1} \bar{S}_i^{(l)}\bar{S}_i^{(l+1)}$ \item
Number of state transitions from busy to free:\\ $N_{01}^i(j) =
\sum\limits_{l=1}^{j-1} (1-\bar{S}_i^{(l)})\bar{S}_i^{(l+1)}$
\end{itemize}
The transition probabilities are estimated:\\
 $ \hat{P}_{01}^i(j) = \frac{N_{01}^i(j)}{N_0^i(j)}$ , $\hat{P}_{11}^i(j) = \frac{N_{11}^i(j)}{N_1^i(j)}$

%

In order to share channel transition probabilities between
secondary transmitter and receiver as dictated by the strategy for
the $L_1=N$ case, values of $N_1^i(j)$, $N_0^i(j)$, $N_{11}^i(j)$
and $N_{01}^i(j)$ for each channel are sent within the transmitted
packet. If $K_{i^*}^{(j)}=1$, the transmitter and receiver update
$\hat{P}_{01}^i(j)$ and $\hat{P}_{11}^i(j)$. Otherwise, the
transmitter only updates $N_1^i(j)$ , $N_0^i(j)$, $N_{11}^i(j)$
and $N_{01}^i(j) $, but uses the old values since the last
successful transmission in order to determine which channel to
access at the beginning of a time slot.

In a nutshell, the proposed algorithm uses the full sensing
capability of the secondary transmitter to decouple the
exploration (i.e., learning) task from the exploitation task.
After an ACK is received, both nodes use the common
observation-based belief vector to make the optimal access
decision. On the other hand, in the absence of the ACK, both nodes
can not use the optimal belief vector in order to maintain
synchronization. In this case, the proposed algorithm opts for a
greedy strategy in order to {\em minimize the time between two
successive} ACKs. At this point, we only conjecture the optimality
of this strategy and continue to work on the proof for the journal
version of this work.

As an analytical benchmark, we have the following upper-bound on
the achievable throughput in this scenario. Assuming that the delayed
side information of all the
primary channels' states $S_i^{(j-1)}$ is given to the secondary
transmitter and receiver, to decide on the channel to access at
time $j$, an upper bound expected throughput per slot is given by:
\begin{equation}\label{R}
    \small{R = \sum\limits_{{S_N}=0}^{1} \cdots \sum\limits_{{S_2}=0}^{1} \sum\limits_{{S_1}=0}^{1}\left[\left(\prod\limits_{i=1}^{N}P_{S_i}\right)\left(\max\limits_{i}\left[P_{S_i1}B_i\right]\right)\right]}
\end{equation}

where, $P_{S_i1} $ denotes the state transition probability for
channel $i$ from state $S_i = (0 , 1)$ to the free state. $P_{S_i}$
is the Markov steady state probability of channel $i$ being free or
busy. The first term in the summation corresponds to the probability
that the $N$ channels are in one of the $2^N$ states, and the second
term represents the highest expected throughput given the current
joint state for the $N$ channels.

A final remark is now in order. Assuming that $P_{11}^i =
P_{01}^i$, a channel's probability of being free, $P_{S_i = 1}$,
becomes independent of the previous state, i.e., $P_{S_i = 1} =
P_{11}^i = P_{01}^i$. In this case, the optimal strategy, assuming
that the transition probabilities are known, is for the secondary
transmitter to access the channel $i^* = \arg \max\limits_{i =
1,\cdots, N} \left[P_{S_i = 1}B_i \right]$ and the expected
throughput becomes $\max\limits_{i = 1,\cdots, N} \left[P_{S_i =
1}B_i\right]$~\cite{Lifeng}. Assuming, however, that the
transition probabilities are unknown but both nodes know that
$P_{11}^i = P_{01}^i$, one can estimate each channel's free
probability $P_{S_i = 1}$ as $\hat{P}_{S_i = 1} = N_1^i(j)/j$. In
Section~\ref{numerical}, we quantify the value of this side
information by comparing the performance of this strategy with our
universal algorithm that does not make any prior assumptions about
the transition probabilities.

\section{The Restless Bandit Scenario: $L_1 = 1$}\label{l11}

Assuming that the transition probabilities are known {\em
a-priori} by the secondary users, the medium access scenario in
this case can be formulated as a partially observable Markov
decision process (POMDP)~\cite{Zhao}. The optimal policy, in this
scenario, must strike a balance between gaining instantaneous
reward by \emph{exploiting} channels based on already known
information, and gaining information for future use by
\emph{exploring} new spectrum opportunities. Motivated by the
prohibitive computational complexity of the optimal strategy, the
authors further proposed a reduced complexity strategy based on
the greedy approach that maximizes the per-slot throughput based
on already known information (exploitation only)~\cite{Zhao}. In a
more recent work~\cite{ZhaoWhitle}, the problem was re-casted as a
restless bandit problem and the Whittle's index approach was used
to construct a more efficient medium access policy~\cite{Whittle}.

Here, we relax the assumption of the {\em a-priori} known
transition probabilities by the secondary transmitter/receiver.
This adds another interesting dimension to the problem since the
blind cognitive MAC protocol must now learn this statistical
information on-line in order to make the appropriate access
decisions. Inspired by previous results of Lai {\em et al.}\ in
the multi-armed bandit setup~\cite{Lifeng}, we propose the
following simple strategy. At the beginning of the $T$ slots, each
of the $N$ primary channels is continuously monitored for an
initial learning period ($LP$) to get an estimate for $P_{11}^i$
and $P_{01}^i$. Then, by assigning Whittle's index $T_i^{(j)}$ to
each channel, we are able to choose which channel to access at
each time slot. In summary, the strategy works as follows.

\begin{enumerate}
  \item \emph{Initial learning period:} Each channel is continuously sensed for $LP$ time slots. At the end of the learning period, the transition probabilities are estimated as $ \hat{P}_{01}^i = \frac{N_{01}^i}{N_0^i}$, $\hat{P}_{11}^i = \frac{N_{11}^i}{N_1^i}$
  \item \emph{Decision:} At the beginning of any time slot ($j > N \times LP$), the secondary transmitter and receiver decide to access channel $i^*(j) = \arg \max\limits_{i = 1,\cdots, N} \left[T_i^{(j)} B_i
  \right]$.
  \item \emph{Sensing:} The secondary transmitter senses channel
  $i^*(j)$.
  \item \emph{Learning:} if $i^*(j) = i^*(j-1)$, update
    $N_{11}^i$, $N_1^i$, $N_{01}^i$, $N_0^i$, $\hat{P}_{11}^i$, and $\hat{P}_{01}^i$.
  \item  \emph{Access:} If $\bar{S}_{i^*}^{(j)} = 1$, the transmitter sends its data packet to the receiver. If the receiver successfully receives a packet, it sends an ACK back to the transmitter.
  \item The transmitter and receiver calculate $\bar{\Omega}^{(j+1)}$ given that:
  \footnotesize
  \begin{equation}\label{4}
    \bar{\omega}_i^{(j+1)} = \begin{cases}
            \bar{P}_{11}^i & \text{if $i(j)=i^*(j) , K_{i^*}^{(j)} = 1$} \\
            \bar{D}_i\bar{P}_{11}^i + \left(1-\bar{D}_i\right)\bar{P}_{01}^i  & \text{if $i(j)=i^*(j) , K_{i^*}^{(j)} = 0$} \\
            \bar{\omega}_i^{(j)}\bar{P}_{11}^i + (1-\bar{\omega}_i^{(j)})\bar{P}_{01}^i & \text{if $i(j) \neq i^*(j)$}\\
           \end{cases}
  \end{equation}
  \normalsize
\end{enumerate}

\noindent where $\bar{P}_{11}^i$ and $\bar{P}_{01}^i$ are the
latest successfully shared $\hat{P}_{11}^i$ and $\hat{P}_{01}^i$
between the secondary transmitter-receiver pair. Finally,
$\bar{\Omega}^{(j+1)}$ is used to update Whittle's index
$T_i^{(j+1)}$ of each channel as detailed in \cite{ZhaoWhitle}.

In the case of time-independent channel states, i.e., $P_{11}^i =
P_{01}^i$, the problem reduces to the a multi-armed bandit
scenario considered in~\cite{Lifeng}. The difference, here, is the
lack of the dedicated control channel, between the cognitive
transmitter and receiver, as assumed in~\cite{Lifeng}. The
following strategy, which is applied as soon as the initial
synchronization is established, avoids this drawback by ensuring
synchronization using the ACK feedback over the same data channel.
\begin{enumerate}
  \item \emph{Decision:} At the beginning of any time slot $j$, the secondary transmitter and receiver decide
  to access the channel $i^*(j) = \arg \max\limits_{i = 1,\cdots, N} \left[\gamma_i^{(j)}
  B_i\right]$, where $\gamma_i^{(j)}=\frac{X_i^{(j)}}{Y_i^{(j)}} +
  \sqrt{\frac{2lnj}{Y_i^{(j)}}}$, $X_i^{(j)}$ is the number of
  time slots where successful communication occurs on channel $i$,
  and $Y_i^{(j)}$ is the number of time slots where channel $i$ is
  chosen to sense and access.
  \item \emph{Sensing:} The secondary transmitter senses channel
  $i^*(j)$.
  \item  \emph{Access:} If $\bar{S}_{i^*}^{(j)} = 1$, the transmitter sends its data packet to the receiver. If the receiver successfully receives a packet, it sends an ACK back to the transmitter.
  \item The transmitter and receiver update the following:\\
        $Y_{i}^{(j+1)} = Y_{i}^{(j)}+1 $, if $i(j)=i^*(j)$\\
        $X_{i}^{(j+1)} = X_{i}^{(j)}+1 $, if $K_{i^*}^{(j)} = 1 , i(j)=i^*(j)$\\
        $\gamma_i^{(j+1)} = \frac{X_i^{(j+1)}}{Y_i^{(j+1)}} + \sqrt{\frac{2lnj}{Y_i^{(j+1)}}}$
\end{enumerate}

\section{Numerical Results}\label{numerical}
\begin{figure}
  \includegraphics[width=.5 \textwidth]{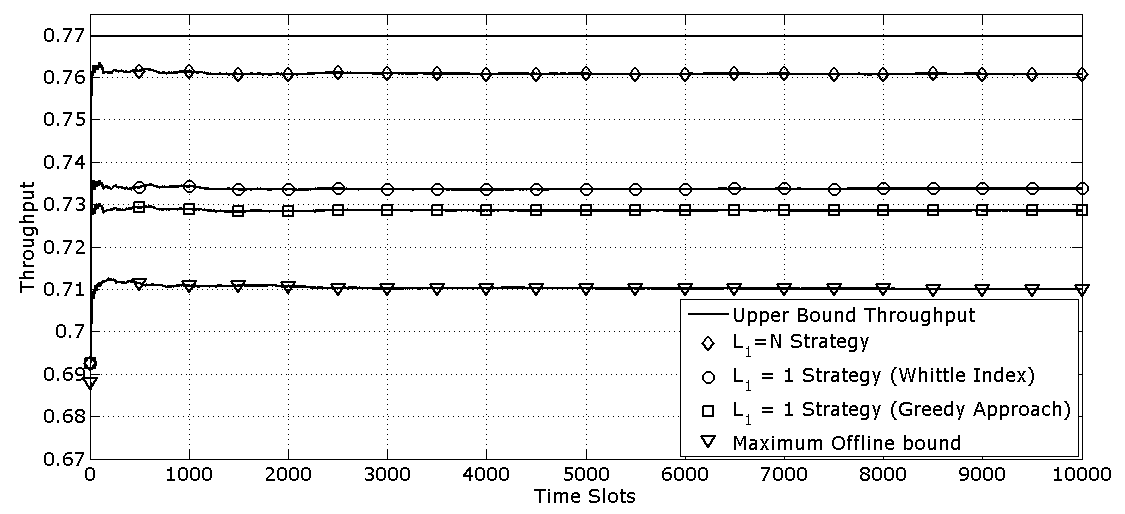}
  \caption{Throughput comparison between: the upper bound from equation [\ref{R}], the proposed blind strategy proposed for $L_1=N$, the Whittle index strategy for $L_1=1$, the greedy strategy for $L_1=1$, and the maximum achievable offline bound.}
\label{fig2}\end{figure}

\begin{figure}
  \includegraphics[width=.5 \textwidth]{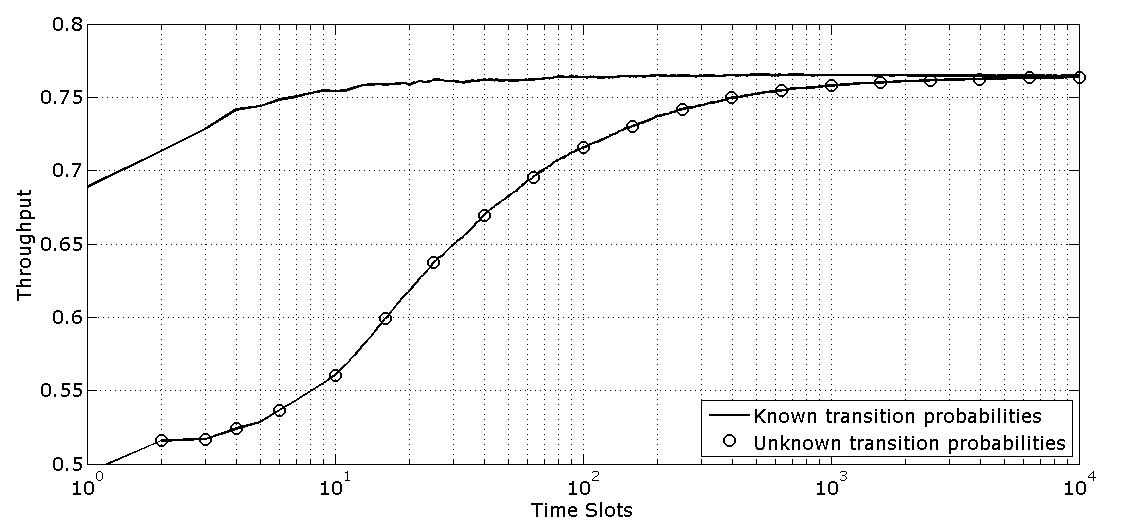}
  \caption{Throughput comparison between the proposed strategy for $(L_1 = N)$ with and without known transition probabilities.}
\label{fig3} \end{figure}

\begin{figure}
  \includegraphics[width=.5 \textwidth]{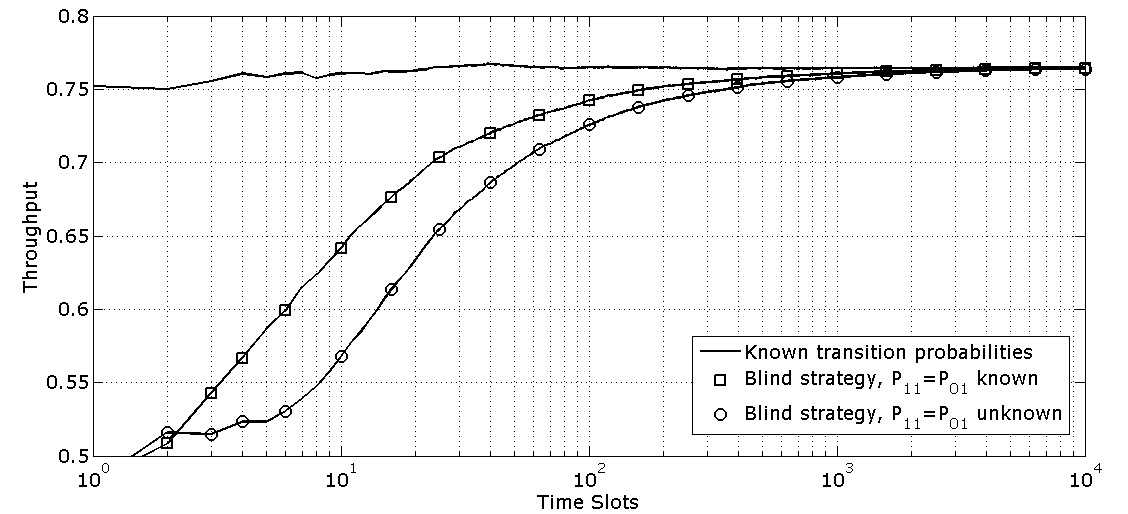}
  \caption{Throughput comparison for the blind cognitive MAC protocol (with and without the prior knowledge that $P_{11}^i = P_{01}^i$) and the genie-aided scenario.}
\label{fig4}\end{figure}

\begin{figure}
  \includegraphics[width=.5 \textwidth]{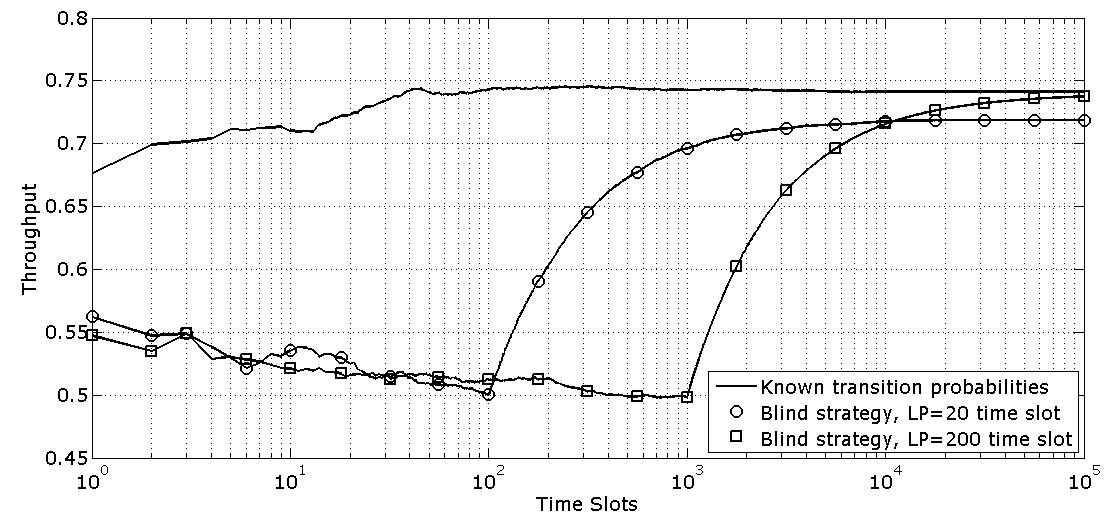}
  \caption{Throughput comparison between the proposed blind strategy for $(L_1 = 1)$, when $LP = 20$ and $LP = 200$, and the genie-aided case.}
\label{fig5}\end{figure}

In this section we present simulation results for the two
scenarios discussed earlier. Throughout this section, we assume
that the number of primary channels $N=5$, each with bandwidth
$B_i = 1$. The spectrum usage statistics of the primary network
were assumed to remain unchanged for a block of $T=10^4$ time
slots for Figures~\ref{fig2},~\ref{fig3}, and~\ref{fig4}, and for
a block of $T=10^5$ time slots for Figure~\ref{fig5}. The transition
probabilities for each channel $P_{11}^i$ and $P_{01}^i$, were
generated randomly between $0.1$ and $0.9$. The plotted results
are the average over $1000$ simulation runs. The discount factor
used to obtain the Whittle index is $0.9999$. In all reported
simulations, perfect sensing is assumed, and the average
throughput per time slot is plotted.

Figure~\ref{fig2} reports the throughput comparison between the
different cognitive MAC strategies, all with prior knowledge about
the channels transition probabilities. The loss in throughput
between the upper bound and the proposed strategy for the $L_1 = N$
case is shown and the gain offered by the full sensing capability as
compared with the $L_1=1$ scenario is apparent. It is seen also that
the strategies we proposed achieve higher throughput than the best
offline bound described in~\cite{Capacity}, in which the channel
with highest steady state probability of being free is always chosen. Figure~\ref{fig3}
illustrates the convergence of the throughput of the proposed blind
strategy for $L_1=N$, with no prior information, to the case with
prior knowledge of the transition probabilities as $T$ grows. In
Figure~\ref{fig4}, we assume that $P_{11}^i = P_{01}^i$ for all
channels. It is shown that even if the secondary users are unaware
of this fact, and apply the proposed strategy, the achievable
throughput converges asymptotically to the achievable performance
when the fact that $P_{11}^i = P_{01}^i$ is known {\em a-priori},
albeit at the expense of a longer learning phase. Interestingly,
both strategies are shown to converge asymptotically to genie-aided
upper bound (when the transition probabilities are known). Finally,
Figure~\ref{fig5} demonstrates the tradeoff between the learning
time overhead in the blind strategy of Section~\ref{l11} and the
final achievable throughput at the end of the $T$ slots. Clearly,
this figure supports the intuitive conclusion that for large $T$
blocks, one can tolerate a longer learning phase in order to
maximize the steady state achievable throughput.

\section{Conclusion}
In this work, we propose blind cognitive MAC protocols that do not
require any prior knowledge about the statistics of the primary
traffic. We differentiate between two distinct scenarios, based on
the complexity of the cognitive transmitter. In the first, the
full sensing capability of the secondary transmitter is fully
utilized to learn the statistics of the primary traffic while
ensuring perfect synchronization between the secondary transmitter
and receiver in the absence of a dedicated control channel. The
second scenario focuses on low-complexity cognitive transmitter
capable of sensing one channel only at the beginning of each time
slot. For this case, we propose an augmented Whittle index MAC
protocol that allows for an initial learning phase to estimate the
transition probabilities of the primary traffic. Our numerical
results demonstrate the convergence of the blind protocols
performance to that of the genie-aided scenario where the primary
traffic statistic are known {\em a-priori} by the secondary
transmitter and receiver.


\begin{thebibliography}{10}

\bibitem{Haykin}
S.~Haykin, "Cognitive radio: brain-empowered wireless communications," \emph{IEEE JSAC},
vol.~23, no.~2, pp.~201-220, Feb~2005.

\bibitem{HangSu}
H.~Su and X.~Zhang, "Opportunistic MAC Protocols for Cognitive Radio," \emph{Proc. 41st Conference on Information
Sciences and Systems (CISS 2007)}, March~2007

\bibitem{HyoilKim}
H.~Kim and K.~Shin, "Efficient Discovery of Spectrum Opportunities with MAC-Layer Sensing in Cognitive Radio Networks," \emph{IEEE Transactions on Mobile Computing},
vol.~7, no.~5, pp.~533-545, May~2008

\bibitem{Capacity}
S.~Srinivasa, S.~Jafar and N.~Jindal, "On the Capacity of the Cognitive Tracking Channel," \emph{IEEE International Symposium on Information Theory}, July~2006

\bibitem{Zhao}
Q.~Zhao, L.~Tong, A.~Swami, and Y.~Chen, "Decentralized Cognitive MAC for Opportunistic Spectrum Access in Ad Hoc Networks: A POMDP Framework," \emph{IEEE JSAC},
vol.~25, no.~3, pp.~589-600, April~2007.

\bibitem{Berkley}
A.~Sahai, N.~Hoven, S.~Mishra and R.~Tandra, "Fundamental tradeoffs in robust spectrum sensing for opportunistic frequency reuse," \emph{Tech. Rep.}, 2006. Available: www.eecs.berkeley.edu/~smm/CognitiveTechReport06.pdf

\bibitem{HMM}
L.~Rabiner and H.~Juang, "An introduction to hidden Markov models," \emph{IEEE ASSP Magazine},
vol.~3, no.~1, Jan.~1986

\bibitem{ZhaoWhitle}
K.~Liu and Q.~Zhao, "A Restless Bandit Formulation of Opportunistic Access: Indexablity and Index Policy," \emph{5th IEEE Annual Communications Society Conference on Sensor, Mesh and Ad Hoc Communications and Networks Workshops (SECON Workshops '08)}, June 2008

\bibitem{Lifeng}
L.~Lai, H.~El-Gamal, H.~Jiang and H.~Poor, "Cognitive Medium Access: Exploration, Exploitation and Competition," \emph{submitted to the IEEE Transactions on Networking}, Oct. 2007

\bibitem{Whittle}
P.~Whittle, "Restless Bandits: Activity Allocation in a Changing World," \emph{Journal of Applied Probability}, vol.~25, 1988

\end{thebibliography}
\end{document}